\documentclass[10pt, twocolumn]{IEEEtran}
\usepackage{graphicx,amsmath,amssymb,cite}
\usepackage{subfigure}
\usepackage{mathabx}
\usepackage{float}
\usepackage{stfloats}
\usepackage{graphicx,subfigure,amsmath,amssymb,amsfonts,cite,float,color,longtable,rotating,diagbox,booktabs,array}
\usepackage{float,color}
\usepackage{algorithm}
\usepackage{algorithmic}
\usepackage{amsmath}

\ifCLASSINFOpdf
\else
\fi

\begin{document}

\title{Secure Hybrid Digital and Analog Precoder for mmWave Systems with low-resolution DACs and finite-quantized phase shifters}

\author{Ling~Xu,~Feng~Shu,~\IEEEmembership{Member,~IEEE},~Guiyang~Xia,\\~Yijin~Zhang,~Zhihong~Zhuang,~and~Jiangzhou~Wang,~\IEEEmembership{Fellow,~IEEE}
\thanks{This work was supported in part by the National Natural Science Foundation of China (Nos. 61771244, 61472190, 61702258, 61501238, and 61602245)(Corresponding authors: Feng Shu,~Ling Xu,~and~Zhihong~Zhuang).}
\thanks{~Ling~Xu,~Feng~Shu,~Guiyang~Xia,~Yijin~Zhang~and Zhihong Zhuang ~are with School of Electronic and Optical Engineering, Nanjing University of Science and Technology, 210094, CHINA. E-mail:\{shufeng, xuling\}@njust.edu.cn}
\thanks{Jiangzhou Wang is with the School of Engineering and Digital Arts, University of Kent, Canterbury CT2 7NT, U.K. Email: \{j.z.wang\}@kent.ac.uk.
}
}

\maketitle
\begin{abstract}
Millimeter wave (mmWave) communication has been regarded as one of the most promising technologies for the future generation wireless networks because of its advantages of providing a ultra-wide new spectrum  and ultra-high data transmission rate. To reduce the power consumption and circuit cost for mmWave systems, hybrid digital and analog (HDA) architecture is preferred in such a scenario. In this paper, an artificial-noise (AN) aided secure HDA beamforming scheme is proposed for mmWave MISO system with low resolution digital-to-analog converters (DACs) and finite-quantized phase shifters on RF. The additive quantization noise model for AN aided HDA system is established to make an analysis of the secrecy performance of such systems. With the partial channel knowledge of eavesdropper available, an approximate expression of secrecy rate (SR) is derived. Then using this approximation formula, we propose a two-layer alternately iterative structure (TLAIS) for optimizing digital precoder (DP) of confidential message (CM), digital AN projection matrix (DANPM) and analog precoder (AP). The inner-layer iteration loop is to design the DP of CMs and DANPM alternatively given a fixed matrix of AP. The outer-layer iteration  loop is in between digital baseband part and analog part, where the former refers to DP and DANPM, and the latter is AP. Then for a given digital part, we propose a gradient ascent algorithm to find  the vector of AP vector.  Given  a matrix of AP, we make use of general power iteration (GPI) method to compute  DP and DANPM. This process is repeated until the terminal condition is reached. Simulation results show that the proposed TLAIS can achieve a better SR performance compared to existing methods, especially in the high signal-to-noise ratio region.
\end{abstract}


%
\IEEEpeerreviewmaketitle

\section{Introduction}
With the rapid development of wireless communication technologies as well as explosive access of mobile terminals, the demand of wireless data traffic grows exponentially. However, the spectrum of current exiting wireless systems is highly congested, which brings a bottleneck for increasing wireless access rate ulteriorly. Millimeter wave (mmWave) communication, whose frequency band ranges from 30GHz to 300GHz, emerges as a promising candidate way of addressing the problem of spectrum congestion \cite{TBai,RW,Guan,WuQ}.

Nevertheless, although the mmWave communication becomes more and more popular due to its abundant available and under-utilized spectrum, its realistic applications are under many constraints such as the severe free space path loss and rain attenuation \cite{TS}. To overcome these challenges, multiple-input-multiple-output (MIMO) is usually adopted to compensate for these shortages. Meanwhile, hybrid digital and analog (HDA) architecture has been employed to further reduce the energy consumption of mmWave MIMO systems \cite{Ayach, YLee, Xianghao, Xinyu, Foad, S.Park, Qin, Sun}. The design of hybrid precoder was formulated as a problem of sparse signal reconstruction by exploiting the inherent spatial sparse structure of mmWave channel in \cite{Ayach, YLee}, where orthogonal matching pursuit (OMP) algorithm was used to achieve a comparatively good performance.
 In \cite{Xianghao}, the authors proposed three methods of innovative alternative minimization for both fully-connected and partially-connected hybrid combiner in mmWave MIMO systems, which confirms the feasibility and effectiveness of hybrid precoder in mmWave MIMO architecture. In \cite{Xinyu}, an energy-efficient successive-interference-cancelation-based algorithm with low complexity was proposed for mmWave MIMO systems. Authors in \cite{Foad} and \cite{S.Park} extended conventional flat-fading mmWave channels to broadband frequency-selective mmWave channels, where the joint analog beamforming for the entire band and respective baseband precoder for each sub-band were required  to be designed carefully. Specifically,  in \cite{Foad} the heuristic algorithms in  OFDM systems  were developed for two scenarios: single-user MIMO and multiple-user multiple-input-single-output scenarios. In \cite{S.Park}, the authors developed a novel method by dynamically establishing the structure of sub-arrays. In \cite{Qin} , the  hybrid analog and digital (HAD) architecture was used  at receiver to make a measurement of direction of arrival (DOA). Here, three low-complexity DOA estimation methods were proposed and the corresponding HAD Cramer-Rao lower bound was also derived.  In \cite{Sun}, a robust beamforming method using HAD receive structure  was  proposed to achieve interference compression.

However, the aforementioned research works mainly focus on HDA mmWave systems  without taking security into consideration. Wyner first proposed a discrete, memoryless wiretap channel model to investigate the secrecy of wireless communication \cite{Wyner}. Based on the framework of Wyner's wiretap channel, multiple-antenna technique \cite{Chen} is applied to enhance security and beamforming has been proven to be secrecy-capacity-achieving under the circumstances that the desired receiver has single antenna and full channel knowledge are available for all terminals\cite{Khistia, Khistib}. If the transmitter can not obtain the simultaneous channel state information (CSI) of eavesdropper or only know the partial CSI of eavesdropper, artificial noise (AN) was shown to be an effective aiding method to strengthen security \cite{Goel, QLi, Negi,Xia}. Recently, some researchers paid a  close attention to security in mmWave MIMO channel. Authors in \cite{Yahia} made a systematical investigation of the security performance in HDA mmWave systems under two different CSIs: full  and partial. Here, maximum ratio transmission (MRT) method was extended to such scenarios and minimum secrecy outage probability was adopted to design the corresponding hybrid precoder. Considering the sparse features of millimeter wave channels in \cite{Y.Ju},   a discrete angular domain channel model was proposed to meticulously derive the secure performance for the proposed transmission schemes in slow fading multipath channels. In \cite{Y.R}, an AN-aided hybrid precoder is proposed to maximize the lower bound of average secrecy rate (SR).

 In practical applications, if  low-resolution digital-to-analog converters (DACs) or analog-to-digital converters (ADCs) are adopted,  the average power consumption of hybrid transceivers will be greatly reduced. Actually, a high -resolution DACs/ADCs can be power-hungry \cite{Orhan}.  In \cite{Abbas}, how to strike a trade-off between energy efficiency and spectrum efficiency was investigated for analog, hybrid, and digital receivers with low resolution ADCs, respectively. To further reduce energy consumption, hybrid beamforming and digital beamforming with low resolution ADCs are studied in \cite{Roth}. Here, the authors find that digital beamforming equipped with finite resolution ADCs can achieve a higher achievable rate and can be more energy efficient than hybrid beamforming in the low signal-to-noise ratio (SNR) region. In \cite{Ribeiro}, taking both low resolution DACs and RF losses in hybrid into account in mmWave MIMO system,  a quantized hybrid transmitter with additive quantization noise model (AQNM) has been constructed for both fully-connected and partially-connected hybrid architecture, and a lower bound of achievable rate is presented.

However, to the best of our knowledge, there are few literature of making an investigation on  how to achieve a security in hybrid mmWave systems with low resolution DACs. Therefore, in our paper, we propose an AN-aided hybrid mmWave transmitter with low resolution DACs and finite-quantized phase shifters. The transmitted signals  with the help of AN are first precoded by digital precoder (DP) in baseband, then passing through low-resolution DACs and RF chains before analog precoder (AP). Here, the AQN model is adopted to approximate the quantized signal as a linear output, which simplifies the analysis of further secrecy rate. Since the optimization problem is non-convex and intractable to tackle,  an alternate iteration algorithm is resorted to maximize the approximate expression of secrecy rate. Our main contributions are summarized as follows:
\begin{enumerate}
 \item An AN-aided secure hybrid precoding system model with low-resolution DACs and finite-quantized phase shifters on RF is established. By taking two kinds of quantization errors (QEs) into consideration, the proposed model is completely distinct from conventional non-secure hybrid precoding system without considering QEs and the AQN quantized model is adopted in our system model to approximate the distortion from QE of low-resolution DACs  as  a linear output. With partial eavesdropping channel knowledge available, an approximate expression for secrecy rate (SR) is derived. This expression converts the original intractable problem into a more easy-to-handle one. Thus, the approximate SR (ASR) expression will significantly simplify the optimization and design of digital precoding (DP) vector of confidential messages, digital AN projection matrix (ANPM)  and analog precoding (AP) vector in what follows.

 \item A two-layer alternatively  iterative structure (TLAIS) is proposed  to maximize the ASR by taking  quantization errors from both low-resolution DAC and phase shifters into consideration. Given AP vector,  the DP vector of confidential messages and  ANPM are alternatively  computed within an interior iterative loop by making use of general power iterative (GPI) method with the aim to maximize the approximate SR. In particular, by complex Kronecker product manipulation, the problem of maximizing ASR with respect to the optimization matrix ANPM is converted into one with respect to an optimization vector being the vectorization of the corresponding ANPM.  Given the DP vector and digital ANPM, a steepest  ascent algorithm is used  to attain the updated AP vector. The above process is repeated until the terminal condition is reached. Finally,  the phases of AP vector is directly taken out as the inputs of the finite quantized phase shifters on RF. More importantly, to reduce the computation complexity, we abstract the non-zero elements in analog part by taking advantage of the sparsity of analog precoder.
\end{enumerate}

The remainder of this paper is organized as follows. Section II presents the system model of AN-aided HDA mmWave multiple input single output (MISO) system with low resolution DACs and finite-quantized phase shifters, and the approximate expression of SR is given in this section. Based on the approximate formula of SR, a TLAIS among DP of confidential message, digital AN projection matrix (DANPM) and AP is proposed in Section III. Performance analysis and simulation evaluations are presented in Section IV. Finally, we draw our conclusions in Section V.

Notation: throughout the paper, matrices, vectors, and scalars are denoted by letters of bold upper case, bold lower case, and lower case, respectively. $(\cdot)^T$, $(\cdot)^*$， and $(\cdot)^H$ denote transpose,~conjugate,~and conjugate transpose,~respectively. $\parallel\cdot\parallel_2$ and $\parallel\cdot\parallel_F$ denote the $l_2$ norm of a vector and Frobenius norm of a matrix, respectively. $\text{Tr}(\cdot)$ and $\mathrm{vec}(\cdot)$ are matrix trace and matrix vectorization; $\otimes$ and $\odot$ indicate the Kronecker products and Hadamard products between two matrices, respectively. $\mathrm{diag}(\mathbf{A})$ returns a diagonal matrix consisting of the corresponding diagonal elements of matrix $\mathbf{A}$. $\mathrm{Diagblk}(\mathbf{a}_1,\mathbf{a}_2,\cdots,\mathbf{a}_N)$ returns the block diagonal matrix concatenated of $\mathbf{a}_1,\mathbf{a}_2,\cdots,\mathbf{a}_N$. And $\mathbb{N}[\mathbf{a}]$ returns a vector consisting of the non-zero elements in $\mathbf{a}$. $\mathbf{A}(m,n)$ denotes the element in $m^{th}$ row and $n^{th}$ column, $\mathbf{A}(m_1:m_2,n_1:n_2)$ returns a matrix consisting of $m_1^{th}$ to $m_2^{th}$ row and $n_1^{th}$ to $n_2^{th}$ column in $\mathbf{A}$.

\newcounter{TempEqCnt} 
\setcounter{TempEqCnt}{\value{equation}} 
\setcounter{equation}{9}
\begin{figure*}
\begin{align}
R_b=\log_2\left(1+\frac{\beta P(1-\eta)^2\mathbf{h}_b\mathbf{F}_{RF}\mathbf{f}_{BB}\mathbf{f}_{BB}^H\mathbf{F}_{RF}^H\mathbf{h}_b^H}{(1-\beta)P(1-\eta)^2\mathbf{h}_b\mathbf{F}_{RF}\mathbf{T}_{BB}\mathbf{T}_{BB}^H\mathbf{F}_{RF}^H\mathbf{h}_b^H+\mathbf{h}_b\mathbf{F}_{RF}\mathbf{R}_{\mathbf{n}_q\mathbf{n}_q}\mathbf{F}_{RF}^H\mathbf{h}_b^H+\sigma^2}\right)
\label{Rd}
\end{align}
\end{figure*}
\setcounter{equation}{\value{TempEqCnt}}

\newcounter{TempEqCnt1} 
\setcounter{TempEqCnt1}{\value{equation}} 
\setcounter{equation}{10}
\begin{figure*}
\begin{align}
R_{e}=\mathbb{E}\left[\log_2\left(1+\frac{\beta P(1-\eta)^2\mathbf{h}_{e}\mathbf{F}_{RF}\mathbf{f}_{BB}\mathbf{f}_{BB}^H\mathbf{F}_{RF}^H\mathbf{h}_{e}^H}{(1-\beta)P(1-\eta)^2\mathbf{h}_{e}\mathbf{F}_{RF}\mathbf{T}_{BB}\mathbf{T}_{BB}^H\mathbf{F}_{RF}^H\mathbf{h}_{e}^H+\mathbf{h}_{e}\mathbf{F}_{RF}\mathbf{R}_{\mathbf{n}_q\mathbf{n}_q}\mathbf{F}_{RF}^H\mathbf{h}_{e}^H+\sigma^2}\right)\right]
\label{Re}
\end{align}
\end{figure*}
\setcounter{equation}{\value{TempEqCnt1}}

\section{Channel and System Models}
\begin{figure}[htp]
\centering
\includegraphics[width=9cm]{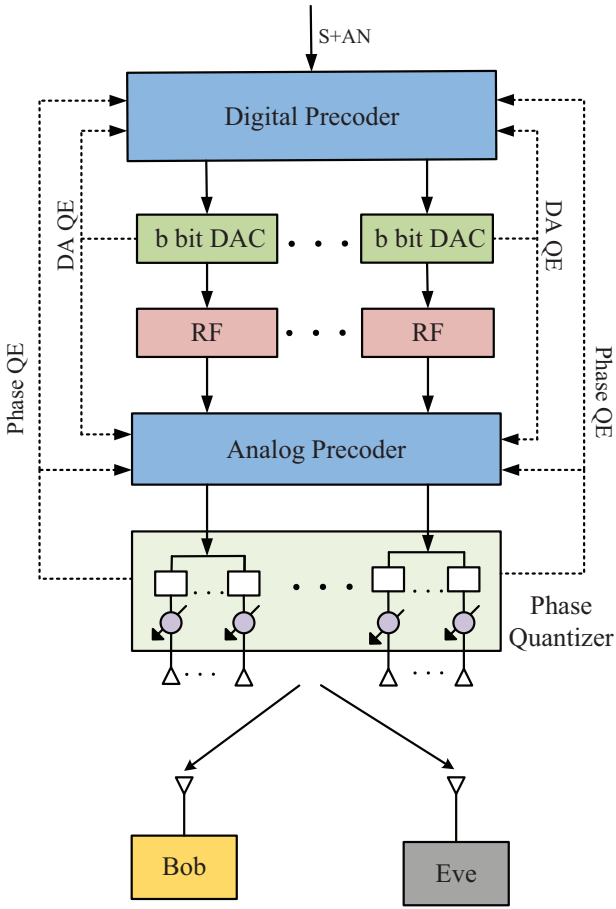}
\caption{Hybrid architecture transmitter with low resolution DACs}
\label{system}
\end{figure}
\subsection{System model}
In this section, we consider a system model with HDA transmitter of low resolution DACs as shown in Fig.~\ref{system}. There are three network nodes: Alice, Bob, and Eve. Working as a transmitter, Alice uses a  partially connected hybrid architecture in this paper,  where each RF chain is connected to a subset of antennas. Assume the transmit antennas at transmitter is $N_T$, and the number of RF chains is $K$, then it is clear that $\mathrm{N_T=KT}$ with each sub-array having $T$ antennas.

The transmit signal can be expressed as follows:
\begin{align}\label{transmit-signal}
\mathbf{x}=\mathbf{F}_{RF}Q_b(\sqrt{\beta P}\mathbf{f}_{BB}s+\sqrt{(1-\beta)P}\mathbf{T}_{BB}\mathbf{z})
\end{align}
where $P$ is the effective transmitted power,~$\beta$ denotes the power allocation (PA) factor of confidential message, and $1-\beta$ indicates the PA factor of AN. $\mathbf{f}_{BB}\in \mathbb{C}^{K\times 1}$ and $\mathbf{T}_{BB}\in \mathbb{C}^{K\times K}$ represent the digital beamforming vector of confidential message and AN projection matrix, respectively. To satisfy the transmit power constraint, we have $\mathbb{E}\{\mathbf{x}\mathbf{x}^H\}=P_T$. $\mathbf{F}_{RF}\in \mathbb{C}^{N\times K}$ is the analog precoding matrix with the following structure:
\begin{align}
\mathbf{F}_{RF}&=\mathrm{Diagblk}(\mathbf{f}_{RF1},\cdots,\mathbf{f}_{RFK})\nonumber\\
&=\left[ {\begin{array}{*{20}{c}}
{\mathbf{f}_{RF1}}&{\mathbf{0}}&{\cdots}&{\mathbf{0}}\\
{\mathbf{0}}&{\mathbf{f}_{RF2}}&{\cdots}&{\mathbf{0}}\\
{\vdots}&{\vdots}&{\ddots}&{\vdots}\\
{\mathbf{0}}&{\mathbf{0}}&{\cdots}&{\mathbf{f}_{RFK}}
\end{array}} \right]
\end{align}
where $\mathbf{f}_{RFk}$ denotes a vector defined as
\begin{align}
\mathbf{f}_{RFk}=[\exp(j\varphi_{k,1}),\exp(j\varphi_{k,2}),\cdots,\exp(j\varphi_{k,M})]^T.
\end{align}
where
\begin{align}
\varphi_{k,i}\in\mathcal{F}_{RF}=\left\{0,\frac{2\pi}{2^{b_{ps}}},\cdots,\frac{2\pi2^{b_{ps}-1}}{2^{b_{ps}}}\right\}
\end{align}
where  $\mathcal{F}_{RF}$ represents the set of $b_{ps}$-bit-quantized phases with $b_{ps}$ being the number of  quantization bits of phase shifters at RF.
In this paper, we apply the additive quantization noise (AQN) model in \cite{Abbas,Ribeiro} to approximate the DACs quantization as a linear output for the simplicity of analysis, which can be formulated as
\begin{align}
Q_b(\mathbf{u})\approx(1-\eta)\mathbf{u}+\mathbf{n}_q
\end{align}
where $\eta$ is defined as the reciprocal of signal-to-quantization-noise ratio. $\mathbf{n}_q$ is the additive quantization noise vector and uncorrelated with the input $\mathbf{u}$, that is, $\mathbb{E}[\mathbf{u}\mathbf{n}_q^H]=\mathbb{E}[\mathbf{n}_q\mathbf{u}^H]=\mathbf{0}$.
Therefore, the transmitted signal in (\ref{transmit-signal}) can be rewritten as
\begin{align}
\mathbf{x}\approx&\sqrt{\beta P}(1-\eta)\mathbf{F}_{RF}\mathbf{f}_{BB}s\nonumber\\
&+\sqrt{(1-\beta)P}(1-\eta)\mathbf{F}_{RF}\mathbf{T}_{BB}\mathbf{z}+\mathbf{F}_{RF}\mathbf{n}_q
\end{align}
According to \cite{Abbas}, the covariance of the quantization noise can be given by
\begin{align}
\mathbf{R}_{\mathbf{n}_q\mathbf{n}_q}=\eta(1-\eta)\mathrm{diag}\{\mathbf{R}_{uu}\}
\end{align}
where $\mathbf{R}_{uu}=\beta P\mathbf{f}_{BB}\mathbf{f}_{BB}^H+(1-\beta)P\mathbf{T}_{BB}\mathbf{T}_{BB}^H$. 
Therefore, the received signal at  Bob can be represented as
\begin{align}
y_b=&\sqrt{\beta P}(1-\eta)\mathbf{h}_b\mathbf{F}_{RF}\mathbf{f}_{BB}s\nonumber\\
&+\sqrt{(1-\beta)P}(1-\eta)\mathbf{h}_b\mathbf{F}_{RF}\mathbf{T}_{BB}\mathbf{z}+\mathbf{h}_b\mathbf{F}_{RF}\mathbf{n}_q+n_b
\end{align}
Similarly, the receive signal at eavesdropper can be given as
\begin{align}
y_{e}&=\sqrt{\beta P}(1-\eta)\mathbf{h}_{e}\mathbf{F}_{RF}\mathbf{f}_{BB}s\nonumber\\
&+\sqrt{(1-\beta)P}(1-\eta)\mathbf{h}_{e}\mathbf{F}_{RF}\mathbf{T}_{BB}\mathbf{z}+\mathbf{h}_{e}\mathbf{F}_{RF}\mathbf{n}_q+n_{e}
\end{align}
where $n_b\sim\mathcal{C}\mathcal{N}(0, \sigma_b^2)$ and $n_e\sim\mathcal{C}\mathcal{N}(0, \sigma_e^2)$ are additive white Gaussian noise (AWGN) at Bob and Eve, respectively. For the convenience of analysis, we set $\sigma_b^2=\sigma_e^2$.

Therefore, the achievable rate at Bob and Eve can be formulated in (\ref{Rd}) and (\ref{Re}), respectively. And we can express the optimization problem of maximizing the SR as
\setcounter{equation}{11}
\begin{align}\label{object_function}
&\max\limits_{\mathbf{F}_{RF}, \mathbf{f}_{BB}, \mathbf{T}_{BB}, \beta}{\kern 1pt}~~~~~~R_s=R_b-R_{e}\nonumber\\
&\text{subject~to}~~~~~~~~~~\mathbf{F}_{RF}\in\mathcal{F}_{RF}, \mathbb{E}\{\mathbf{x}\mathbf{x}^H\}=P_T, 0\!\leq\beta\leq\!1
\end{align}
Since it is assumed that only the partial knowledge of wiretap channel at Eve is available, we can only obtain the average achievable rate of Eve. Meanwhile, because of partially-connected structure, we have $\mathbf{F}_{RF}^H\mathbf{F}_{RF}=\mathbf{I}_K$. To obey the power constraint, we should have $\|\mathbf{f}_{BB}\|^2=\|\mathbf{T}_{BB}\|_F^2=1$, and $P=\frac{P_T}{(1-\eta)^2+\eta(1-\eta)}$. Therefore, the objective function in  optimization problem (\ref{object_function}) can be lower bounded as:
\begin{align}\label{lower_of}
R_s&=R_b-\mathbb{E}[\log_2\left\{1+\mathrm{SINR}_e\right\}] \nonumber\\
&\overset{\text{(a)}}\geq R_b-\log_2\left\{1+\mathbb{E}[\mathrm{SINR}_e]\right\}\nonumber\\
&=\tilde{R}_s
\end{align}
where $(a)$ holds due to
\begin{align}
\mathbb{E}[\log_2(x)]\leq\log_2(1+\mathbb{E}[x]).
\end{align}
Therefore, replacing the objective function  $R_s$ in (\ref{object_function}) by  $\tilde{R}_s$ forms the following simplified optimization problem
\begin{align}\label{Rs_low}
&\max\limits_{\beta, \mathbf{f}_{BB}, \mathbf{T}_{BB}, \mathbf{F}_{RF}}{\kern 1pt}~~~~~~\tilde{R}_s=R_b-\log_2\left\{1+\mathbb{E}[\mathrm{SINR}_e]\right\}\nonumber\\
&\text{subject~to}~~~\mathbf{F}_{RF}\in\mathcal{F}_{RF}, \|\mathbf{f}_{BB}\|^2=1,\|\mathbf{T}_{BB}\|_F^2=1, 0\!\leq\beta\leq\!1
\end{align}

\subsection{Channel model}
In what follows,  a narrow-band clustered mmWave channel model is used with $L$ propagation paths, which can be described as follows:
\begin{align}\label{channel}
\mathbf{h}=\sqrt{\frac{N_t}{L}}\sum_{l=1}^Lg_l\mathbf{a}_{t,l}^H(\phi_l)
\end{align}
where $\sqrt{\frac{N_t}{L}}$ is the normalized factor, $l$ stands for the index of paths,  and $L$  is the number of channel paths. The path gains $g_l\sim\mathcal{C}\mathcal{N}(0,1)$ depicts the complex gain of the $l^{th}$ path. $\mathbf{a}_t(\phi_l)$ is the corresponding response vector of transmit antenna array, with $\phi_l$ denotes the azimuth angles, respectively. For uniform linear array with $N$ elements, the array response vector can be represented as
\begin{align}
\mathbf{a}(\phi)=\frac{1}{\sqrt{N}}[1,e^{-j\frac{2\pi}{\lambda }dsin(\phi)},\cdots,e^{-j\frac{2\pi}{\lambda}(N-1)dsin(\phi)}]^T
\end{align}
where $\lambda$ and $d$ are the wavelength of the signal and distance spacing between the antenna elements.

Since we consider the mmWave channel with partial channel knowledge of Eve, i.e., the full knowledge of angles of departure (AoD) and the distribution of the gains of mmWave paths $\alpha_l$. As we can see from (\ref{channel}), the channel can be written more compactly as
\begin{align}
\mathbf{h}=\sqrt{\frac{N_t}{L}}\mathbf{g}\mathbf{A}_t
\end{align}
with $\mathbf{g}=[g_1,g_2,\cdots,g_L]$ and $\mathbf{A}_t=[\mathbf{a}_{t,1}^H,\mathbf{a}_{t,2}^H,\cdots,\mathbf{a}_{t,L}^H]^T$. Since the each element in $\mathbf{g}$ follows the complex Gaussian distribution with zero mean and unit variance, $\mathbf{g}$ is the i.i.d. complex Gaussian vector with probability density function as
\begin{align}
f(\mathbf{g})=\frac{1}{\pi^L}e^{-\mathbf{g}^H\mathbf{g}}.
\end{align}
Therefore, the expectation $\mathbb{E}\{\mathrm{SINR}_{e}\}$ can be approximated as (\ref{ESINR_em}). The approximation in $(b)$ is similarly adopted in \cite{PLin, WLiao, XZhou}.

\section{Design and
Optimization of DP， AP，and ANPM }
\begin{figure}[htp]
\centering
\includegraphics[width=8cm]{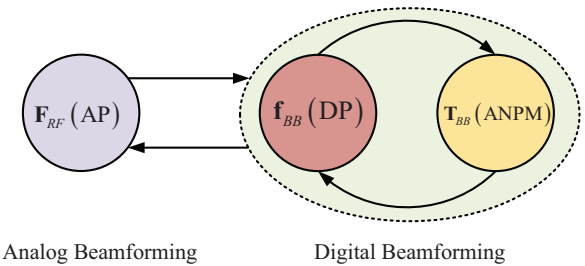}
\caption{Schematic diagram of TLAIS}
\label{iterFig}
\end{figure}

In this section, we propose a TLAIS of maximizing the ASR by taking  quantization errors from both low-resolution DAC and phase shifters into account as shown in Fig.~\ref{iterFig}. For the outer loop, given the initial DP vector of confidential messages and  ANPM, a steepest  ascent  algorithm is used  to attain the updated AP vector by maximizing the approximate SR (ASR).  After completing this, the optimization process turns to the inner loop. Given the AP matrix, the DP vector of confidential messages and  ANPM are alternatively  attained within an interior iterative loop by using GPI method with the aim to maximize the ASR. The two loops are repeated until their individual terminal conditions are satisfied.
%

\newcounter{TempEqCnt2} 
\setcounter{TempEqCnt2}{\value{equation}} 
\setcounter{equation}{19}
\begin{figure*}
\begin{align}
&\mathbb{E}\{\mathrm{SINR}_{e}\}=\mathbb{E}\left\{\frac{\beta P(1-\eta)^2\frac{N_t}{L}\mathbf{g}_{e}\mathbf{A}_{te}\mathbf{F}_{RF}\mathbf{f}_{BB}\mathbf{f}_{BB}^H\mathbf{F}_{RF}^H\mathbf{A}_{te}^H\mathbf{g}_{e}^H}{(1-\beta)P(1-\eta)^2\frac{N_t}{L}\mathbf{g}_{e}\mathbf{A}_{te}\mathbf{F}_{RF}\mathbf{T}_{BB}\mathbf{T}_{BB}^H\mathbf{F}_{RF}^H\mathbf{A}_{te}^H\mathbf{g}_{e}^H+\frac{N_t}{L}\mathbf{g}_{e}\mathbf{A}_{te,m}\mathbf{F}_{RF}\mathbf{R}_{\mathbf{n}_q\mathbf{n}_q}\mathbf{F}_{RF}^H\mathbf{A}_{te}^H\mathbf{g}_{e}^H+\sigma^2}\right\}\nonumber\\
&\overset{\text{(b)}}{\approx}\frac{\mathbb{E}\left\{\beta P(1-\eta)^2\frac{N_t}{L}\mathbf{g}_{e}\mathbf{A}_{te}\mathbf{F}_{RF}\mathbf{f}_{BB}\mathbf{f}_{BB}^H\mathbf{F}_{RF}^H\mathbf{A}_{te}^H\mathbf{g}_{e}^H\right\}}{\mathbb{E}\left\{(1-\beta)P(1-\eta)^2\frac{N_t}{L}\mathbf{g}_{e}\mathbf{A}_{te}\mathbf{F}_{RF}\mathbf{T}_{BB}\mathbf{T}_{BB}^H\mathbf{F}_{RF}^H\mathbf{A}_{te}^H\mathbf{g}_{e}^H+\frac{N_t}{L}\mathbf{g}_{e}\mathbf{A}_{te}\mathbf{F}_{RF}\mathbf{R}_{\mathbf{n}_q\mathbf{n}_q}\mathbf{F}_{RF}^H\mathbf{A}_{te}^H\mathbf{g}_{e}^H+\sigma^2\right\}}\nonumber\\
&=\frac{\beta P(1-\eta)^2\frac{N_t}{L}\mathbb{E}\left\{\mathrm{tr}[\mathbf{g}_{e}^H\mathbf{g}_{e,m}\mathbf{A}_{te,m}\mathbf{F}_{RF}\mathbf{f}_{BB}\mathbf{f}_{BB}^H\mathbf{F}_{RF}^H\mathbf{A}_{te}^H]\right\}}{(1-\beta)P(1-\eta)^2\frac{N_t}{L}\mathbb{E}\left\{\mathrm{Tr}[\mathbf{g}_{e}^H\mathbf{g}_{e}\mathbf{A}_{te}\mathbf{F}_{RF}\mathbf{T}_{BB}\mathbf{T}_{BB}^H\mathbf{F}_{RF}^H\mathbf{A}_{te}^H]+\frac{N_t}{L}\mathrm{Tr}[\mathbf{g}_{e}^H\mathbf{g}_{e}\mathbf{A}_{te}\mathbf{F}_{RF}\mathbf{R}_{\mathbf{n}_q\mathbf{n}_q}\mathbf{F}_{RF}^H\mathbf{A}_{te}^H]+\sigma^2\right\}}\nonumber\\
&=\frac{\beta P(1-\eta)^2\frac{N_t}{L}\mathrm{tr}[\mathbb{E}\left\{\mathbf{g}_{e}^H\mathbf{g}_{e,m}\right\}\mathbf{A}_{te,m}\mathbf{F}_{RF}\mathbf{f}_{BB}\mathbf{f}_{BB}^H\mathbf{F}_{RF}^H\mathbf{A}_{te}^H]}{(1-\beta)P(1-\eta)^2\frac{N_t}{L}\mathrm{tr}[\mathbb{E}\left\{\mathbf{g}_{e}^H\mathbf{g}_{e}\right\}\mathbf{A}_{te}\mathbf{F}_{RF}\mathbf{T}_{BB}\mathbf{T}_{BB}^H\mathbf{F}_{RF}^H\mathbf{A}_{te}^H]+\frac{N_t}{L}\mathrm{Tr}[\mathbb{E}\left\{\mathbf{g}_{e}^H\mathbf{g}_{e}\right\}\mathbf{A}_{te}\mathbf{F}_{RF}\mathbf{R}_{\mathbf{n}_q\mathbf{n}_q}\mathbf{F}_{RF}^H\mathbf{A}_{te}^H]+\sigma^2}\nonumber\\
&=\frac{\beta P(1-\eta)^2\frac{N_t}{L}\mathrm{tr}[\mathbf{A}_{te}\mathbf{F}_{RF}\mathbf{f}_{BB}\mathbf{f}_{BB}^H\mathbf{F}_{RF}^H\mathbf{A}_{te,m}^H]}{(1-\beta)P(1-\eta)^2\frac{N_t}{L}\mathrm{Tr}[\mathbf{A}_{te}\mathbf{F}_{RF}\mathbf{T}_{BB}\mathbf{T}_{BB}^H\mathbf{F}_{RF}^H\mathbf{A}_{te}^H]+\frac{N_t}{L}\mathrm{Tr}[\mathbf{A}_{te}\mathbf{F}_{RF}\mathbf{R}_{\mathbf{n}_q\mathbf{n}_q}\mathbf{F}_{RF}^H\mathbf{A}_{te}^H]+\sigma^2}
\label{ESINR_em}
\end{align}
\end{figure*}
\setcounter{equation}{\value{TempEqCnt2}}

\newcounter{TempEqCnt4} 
\setcounter{TempEqCnt4}{\value{equation}} 
\setcounter{equation}{29}
\begin{figure*}\hrulefill
\begin{align}\label{SINRb}
\mathrm{SINR}_b&=\frac{\beta P(1-\eta)^2\mathbf{h}_b\mathbf{F}_{RF}\mathbf{f}_{BB}\mathbf{f}_{BB}^H\mathbf{F}_{RF}^H\mathbf{h}_b^H}{(1-\beta)P(1-\eta)^2\mathbf{h}_b\mathbf{F}_{RF}\mathbf{T}_{BB}\mathbf{T}_{BB}^H\mathbf{F}_{RF}^H\mathbf{h}_b^H+\mathbf{h}_b\mathbf{F}_{RF}\mathbf{R}_{\mathbf{n}_q\mathbf{n}_q}\mathbf{F}_{RF}^H\mathbf{h}_b^H+\sigma^2}\nonumber\\
&=\frac{\beta P(1-\eta)^2\mathrm{tr}[\mathbf{F}_{RF}^H\mathbf{h}_b^H\mathbf{h}_b\mathbf{F}_{RF}\mathbf{f}_{BB}\mathbf{f}_{BB}^H]}{(1-\beta)P(1-\eta)^2\mathrm{tr}[\mathbf{F}_{RF}^H\mathbf{h}_b^H\mathbf{h}_b\mathbf{F}_{RF}\mathbf{T}_{BB}\mathbf{T}_{BB}^H]+\mathrm{tr}[\mathbf{F}_{RF}^H\mathbf{h}_b^H\mathbf{h}_b\mathbf{F}_{RF}\mathbf{R}_{\mathbf{n}_q\mathbf{n}_q}]+\sigma^2}\nonumber\\
&=\frac{\beta P(1-\eta)^2\mathrm{vec}(\mathbf{F}_{RF})^H[\mathbf{F}_{BB}\otimes\mathbf{H}_b]\mathrm{vec}(\mathbf{F}_{RF})}{(1-\beta)P(1-\eta)^2\mathrm{vec}(\mathbf{F}_{RF})^H[(\mathbf{T}_{BB}\mathbf{T}_{BB}^H)^T\otimes\mathbf{H}_b]\mathrm{vec}(\mathbf{F}_{RF})+\mathrm{vec}(\mathbf{F}_{RF})^H[\mathbf{R}_{nqnq}^T\otimes\mathbf{H}_b]\mathrm{vec}(\mathbf{F}_{RF})+\sigma^2}\nonumber\\
&=\frac{\mathbf{d}^H\mathbf{\Gamma}_{b1}\mathbf{d}}{\mathbf{d}^H[\mathbf{\Gamma}_{b2}+\mathbf{\Gamma}_{b3}+\frac{\sigma^2}{K}\mathbf{I}_Nt]\mathbf{d}}
\end{align}
\end{figure*}
\setcounter{equation}{\value{TempEqCnt4}}

\subsection{Design of AP matrix}
Observing the structure of the numerator  in (\ref{ESINR_em}), we can simplify it to  the following expression
\setcounter{equation}{20}
\begin{align}\label{numerator}
&~~~\mathrm{Tr}[\mathbf{A}_{te,m}\mathbf{F}_{RF}\mathbf{f}_{BB}\mathbf{f}_{BB}^H\mathbf{F}_{RF}^H\mathbf{A}_{te,m}^H]\nonumber\\
&=\mathrm{Tr}[\mathbf{F}_{RF}^H\mathbf{A}_{te,m}^H\mathbf{A}_{te,m}\mathbf{F}_{RF}\mathbf{f}_{BB}\mathbf{f}_{BB}^H]\nonumber\\
&\overset{\text{(c)}}=\mathrm{vec}(\mathbf{F}_{RF})^H[(\mathbf{f}_{BB}\mathbf{f}_{BB}^H)^T\otimes(\mathbf{A}_{te,m}^H\mathbf{A}_{te,m})]\mathrm{vec}(\mathbf{F}_{RF})\nonumber\\
&=\mathrm{vec}(\mathbf{F}_{RF})^H[\mathbf{F}_{BB}\otimes\mathbf{A}]\mathrm{vec}(\mathbf{F}_{RF})
\end{align}
where $\mathbf{F}_{BB}=(\mathbf{f}_{BB}\mathbf{f}_{BB}^H)^T\in\mathbb{C}^{K\times K}$ and $\mathbf{A}=\mathbf{A}_{te,m}^H\mathbf{A}_{te,m}\in\mathbb{C}^{N_T\times N_T}$. In the above equation,  $(c)$ holds due to the fact that
\begin{align}
\mathrm{Tr}\left(\mathbf{P}^H\mathbf{Z}\mathbf{P}\mathbf{W}\right)=\mathrm{vec}(\mathbf{P})^H\left(\mathbf{W}^T\otimes\mathbf{Z}\right)\mathrm{vec}(\mathbf{P}).
\end{align}
Since the analog beamforming matrix $\mathbf{F}_{RF}$ (\ref{numerator}) is block diagonal and most of its elements  are zeros, i.e., the dimension of $\mathbf{F}_{RF}$ is $N_T\times K$. In other words, there are $N_T$ non-zero elements and $(K-1)N_T$ zero elements for  matrix $\mathbf{F}_{RF}$, so the matrix $\mathbf{F}_{RF}$ can be viewed as a sparse matrix. In order to reduce the computational complexity and dimension in the following,  only non-zero elements of $\mathbf{F}_{RF}$ are required to be extracted. Because there are  $N_T$ non-zero elements in $\mathbf{F}_{RF}$,  let us define
\begin{align}
\mathbf{d}=\mathbb{N}[\mathrm{vec}(\mathbf{F}_{RF})]=[\mathbf{f}_{RF1},\mathbf{f}_{RF2},\cdots,\mathbf{f}_{RFK}]^T\in\mathbb{C}^{N\times 1}
\end{align}
represent the mathematic operation of extracting all non-zero elements from $\mathbf{F}_{RF}$. To  match the above operation, now,  let us extract extract the corresponding $N\times N$ elements from $NK\times NK$ matrix $\left(\mathbf{F}_{BB}\otimes\mathbf{A}\right)$ in (\ref{numerator}). The rule of extracting method is as follows: matrix $\mathbf{F}_{BB}\otimes\mathbf{A}$ can be partitioned into $K\times K$ or $K^2$ block matrices with each  being $N\times N$ submatrix,

each block matrix can be expressed as $\mathbf{F}_{BB}(m,n)\otimes\mathbf{A}$ with $m=1,\cdots,K, n=1,\cdots,K$. Only $M\times M$ elements per block matrix  are extracted  with the following rule:
\begin{align}
&\mathbf{\Gamma}\left((m-1)M+1:mM,(n-1)M+1:nM\right)\nonumber\\
=&[\mathbf{F}_{BB}(m,n)\otimes\mathbf{A}]\left(\!(m-1)\!M+1\!:\!mM,\!(n-1)\!M+1\!:\!nM\right)
\end{align}
Via the above manipulation, we have a simple form of (\ref{numerator}) as follows
\begin{align}
\mathrm{vec}(\mathbf{F}_{RF})^H[\mathbf{F}_{BB}\otimes\mathbf{A}]\mathrm{vec}(\mathbf{F}_{RF})=\mathbf{d}^H\mathbf{\Gamma}\mathbf{d}
\end{align}
In the same fashion, (\ref{ESINR_em}) can be rewritten in a more concise way as follows:
\begin{align}
\mathbb{E}\{\mathrm{SINR}_{e}\}&\approx\frac{\mathbf{d}^H\mathbf{\Gamma}_{e1}\mathbf{d}}{\mathbf{d}^H\mathbf{\Gamma}_{e2}\mathbf{d}+\mathbf{d}^H\mathbf{\Gamma}_{e3}\mathbf{d}+\sigma^2}\nonumber\\
&=\frac{\mathbf{d}^H\mathbf{\Gamma}_{e1}\mathbf{d}}{\mathbf{d}^H[\mathbf{\Gamma}_{e2}+\mathbf{\Gamma}_{e3}+\frac{\sigma^2}{K}\mathbf{I}_{Nt}]\mathbf{d}}
\end{align}
Similarly,  the achievable rate of Eve is also rewritten as
\begin{align}
\log_2\left(1+\mathbb{E}[\mathrm{SINR}_e]\right)\approx\log_2\left(\frac{\mathbf{d}^H\mathbf{A}_{e,1}\mathbf{d}}{\mathbf{d}^H\mathbf{A}_{e,2}\mathbf{d}}\right)
\end{align}
with
\begin{align}
&\mathbf{A}_{e,1}=\mathbf{\Gamma}_{e1}+\mathbf{\Gamma}_{e2}+\mathbf{\Gamma}_{e3}+\frac{\sigma^2}{K}\mathbf{I}_{Nt},
\end{align}
and
\begin{align}
&\mathbf{A}_{e,2}=\mathbf{\Gamma}_{e2}+\mathbf{\Gamma}_{e3}+\frac{\sigma^2}{K}\mathbf{I}_{Nt}.
\end{align}

Similarly, the $\mathrm{SINR}_b$ can be also expressed as (\ref{SINRb}). Therefore, we can further write the achievable rate of Bob as
\setcounter{equation}{30}
\begin{align}
R_b=\log_2\left(\frac{\mathbf{d}^H\mathbf{A}_{b,1}\mathbf{d}}{\mathbf{d}^H\mathbf{A}_{b,2}\mathbf{d}}\right)
\end{align}
with
\begin{align}
\mathbf{A}_{b,1}=\mathbf{\Gamma}_{b1}+\mathbf{\Gamma}_{b2}+\mathbf{\Gamma}_{b3}+\frac{\sigma^2}{K}\mathbf{I}_{Nt}\nonumber\\ \end{align}
and
\begin{align}
\mathbf{A}_{b,2}=\mathbf{\Gamma}_{b2}+\mathbf{\Gamma}_{b3}+\frac{\sigma^2}{K}\mathbf{I}_{Nt}.
\end{align}
Finally, the optimization problem of maximizing $\tilde{R}_s$ (\ref{Rs_low}) can be further recasted as
\begin{align}
\tilde{R}_s&=R_b-\log_2\left(1+\mathbb{E}[\mathrm{SINR}]\right)\nonumber\\
&\approx\log_2\left(\frac{\mathbf{d}^H\mathbf{A}_{b,1}\mathbf{d}}{\mathbf{d}^H\mathbf{A}_{b,2}\mathbf{d}}\cdot\frac{\mathbf{d}^H\mathbf{A}_{e,2}\mathbf{d}}{\mathbf{d}^H\mathbf{A}_{e,1}\mathbf{d}}\right)
\end{align}
Since the $\tilde{R}_s$ is the non-convex function of $\mathbf{d}$. In particular, all elements in $\mathbf{d}$ have the unit modulus, which satisfies $\mathbf{d}\in\mathcal{F}_{RF}$. Therefore,  a gradient ascent (GA) method is used to compute the AP matrix. Let us define
\begin{align}
f(\mathbf{d})=\frac{\mathbf{d}^H\mathbf{A}_{b,1}\mathbf{d}}{\mathbf{d}^H\mathbf{A}_{b,2}\mathbf{d}},
g(\mathbf{d})=\frac{\mathbf{d}^H\mathbf{A}_{e,2}\mathbf{d}}{\mathbf{d}^H\mathbf{A}_{e,1}\mathbf{d}}.
\end{align}
The gradient of  $\tilde{R}_s$  with respect to $\mathbf{d}$ in can be given by
\begin{align}\label{gradient_d}
&\nabla_\mathbf{d}=\frac{(f^{'}(\mathbf{d})g(\mathbf{d})+f(\mathbf{d})g^{'}(\mathbf{d}))}{f(\mathbf{d})g(\mathbf{d})\mathrm{ln}2}
\end{align}
where
\begin{subequations}
\begin{align}
&f^{'}(\mathbf{d})=\frac{\mathbf{A}_{b,1}^H\mathbf{d}(\mathbf{d}^H\mathbf{A}_{b,2}\mathbf{d})-(\mathbf{d}^H\mathbf{A}_{b,1}\mathbf{d})\mathbf{A}_{b,2}^H\mathbf{d}}{(\mathbf{d}^H\mathbf{A}_{b,2}\mathbf{d})^2},\\
&g^{'}(\mathbf{d})=\frac{\mathbf{A}_{e,2}^H\mathbf{d}(\mathbf{d}^H\mathbf{A}_{e,1}\mathbf{d})-(\mathbf{d}^H\mathbf{A}_{e,2}\mathbf{d})\mathbf{A}_{e,1}^H\mathbf{d}}{(\mathbf{d}^H\mathbf{A}_{e,1}\mathbf{d})^2}.
\end{align}
\end{subequations}
After obtaining the $\nabla_{\mathbf{d}}$, we will renew the value $\mathbf{d}^{(t)}$ of $\mathbf{d}$ by $\mathbf{d}^{(t-1)}+\alpha\nabla_\mathbf{d}$ with $\alpha$ being the searching step. The detailed process of GA algorithm  proposed by us is listed in Algorithm 1.

\begin{algorithm}[htb]
\caption{Gradient ascent algorithm for analog precoder}
\textbf{Input}:\\
(1) Initialize $\mathbf{F}_{RF}^{(0)}$ according to (\ref{Frf_initial}) and extract non-zero elements in $\mathbf{F}_{RF}^{(0)}$ to initialize $\mathbf{d}^{(0)}$;\\
(2) Fixed $\mathbf{f}_{BB}$ and $\mathbf{T}_{BB}$, compute $R_s^{(0)}$ and set $t=1, \alpha$, threshold value $\alpha_{min}, \epsilon$;\\
~\textbf{for} $\alpha>\alpha_{min}$
\begin{algorithmic}[1] 
\STATE Compute $\nabla_{\mathbf{d}}^{(t-1)}$ according to (\ref{gradient_d}) and $\mathbf{d}^{(t)}=\mathbf{d}^{(t-1)}+\alpha\nabla_\mathbf{d}^{(t-1)}$, reform $\mathbf{F}_{RF}^{(t)}=\frac{1}{\sqrt{M}}\exp\{j\angle(\mathbf{d}^{(t)})\}$;\\
\STATE Compute $R_s^{(t)}$ using $\mathbf{T}_{BB}$, $\mathbf{f}_{BB}$ and $\mathbf{F}_{RF}^{(t)}$, search the optimal $\beta^{t}$ by 1-D search;\\
\STATE \textbf{If} $R_s^{(t)}-R_s^{(t-1)}>\epsilon$\\
~~~~$\mathbf{d}^{(t)}=\mathbf{d}^{(t-1)}+\alpha\nabla_\mathbf{d}^{(t-1)}$;\\
\textbf{else}\\
~~~~$\mathbf{d}^{(t)}=\mathbf{d}^{(t-1)}$;\\
~~~~$\alpha=\frac{\alpha}{2}$;\\
\STATE $t=t+1$;\\
\end{algorithmic}
~\textbf{end for}\\
\textbf{Output}: $\mathbf{F}_{RF}^{(t)}$.
\end{algorithm}

\subsection{Design of DP vector of confidential message}
In this section, we will design the DP vector $\mathbf{f}_{BB}$ assuming the other two precoders  $\mathbf{F}_{RF}$ and $\mathbf{T}_{BB}$ are known in advance. To optimize  $\mathbf{f}_{BB}$, the achievable rate of Bob is represented as a function of $\mathbf{f}_{BB}$ as follows
\begin{align}\label{Rb-f-BB}
R_b&=\log_2\left(1+\frac{\xi_1\mathbf{f}_{BB}^H\mathbf{F}_{RF}^H\mathbf{h}_b^H\mathbf{h}_b\mathbf{F}_{RF}\mathbf{f}_{BB}}{\xi_2\mathbf{h}_b\mathbf{F}_{RF}\mathrm{diag}[\mathbf{f}_{BB}\mathbf{f}_{BB}^H]\mathbf{F}_{RF}^H\mathbf{h}_b^H+\gamma_b}\right)\nonumber\\
&\overset{\text{(d)}}=\log_2\left(1+\frac{\xi_1\mathbf{f}_{BB}^H\mathbf{F}_{RF}^H\mathbf{h}_b^H\mathbf{h}_b\mathbf{F}_{RF}\mathbf{f}_{BB}}{\xi_2\mathbf{h}_b\mathbf{F}_{RF}[(\mathbf{f}_{BB}\mathbf{f}_{BB}^H)\odot\mathbf{I}_K]\mathbf{F}_{RF}^H\mathbf{h}_b^H+\gamma_b}\right)\nonumber\\
&\overset{\text{(e)}}=\log_2\left(1+\frac{\xi_1\mathbf{f}_{BB}^H\tilde{\mathbf{h}}_b^H\tilde{\mathbf{h}}_b\mathbf{f}_{BB}}{\xi_2\mathbf{f}_{BB}^H[\mathbf{I}_K\odot(\tilde{\mathbf{h}}_b^H\tilde{\mathbf{h}}_b)]\mathbf{f}_{BB}+\gamma_b}\right)\nonumber\\
&=\log_2\left(\frac{\mathbf{f}_{BB}^H\mathbf{Q}_b\mathbf{f}_{BB}}{\mathbf{f}_{BB}^H\mathbf{P}_b\mathbf{f}_{BB}}\right)
\end{align}

where
\begin{align}
\xi_1=\beta P(1-\eta)^2
\end{align}
\begin{align}
 \xi_2=\eta(1-\eta)\beta P,
\end{align}
\begin{align}\label{}
\gamma_b&=(1-\beta)P(1-\eta)^2\|\mathbf{h}_b\mathbf{F}_{RF}\mathbf{T}_{BB}\|^2\nonumber\\
&+\eta(1-\eta)(1-\beta)P
\end{align}
\begin{align}
\tilde{\mathbf{h}}_b=\mathbf{h}_b\mathbf{F}_{RF},
\end{align}
\begin{align}
\mathbf{Q}_b=\xi_2[\mathbf{I}_K\odot(\tilde{\mathbf{h}}_b^H\tilde{\mathbf{h}}_b)]+\gamma_b\mathbf{I}_K+\xi_1\tilde{\mathbf{h}}_b^H\tilde{\mathbf{h}}_b,
\end{align}
and
\begin{align}
\mathbf{P}_b=\xi_2[\mathbf{I}_K\odot(\tilde{\mathbf{h}}_b^H\tilde{\mathbf{h}}_b)]+\gamma_b\mathbf{I}_K.
\end{align}
Equality $(d)$ in (\ref{Rb-f-BB})  is achieved because $\mathrm{diag}[\mathbf{A}]=\mathbf{A}\odot\mathbf{I}$ and $(e)$ holds due to the fact that  $\mathrm{Tr}[(\mathbf{A}\odot\mathbf{I})\mathbf{B}]=\mathrm{Tr}[\mathbf{A}(\mathbf{I}\odot\mathbf{B})]$.

 Now, we can rewrite $\mathbb{E}\{\mathrm{SINR}_{e}\}$ as a function of $\mathbf{f}_{BB}$ from (\ref{ESINR_em}) as follows:
\begin{align}
&\mathop{\mathbb{E}[\mathrm{SINR}_{e}]}\limits_{\mathbf{f}_{BB}}\approx\frac{\lambda_1\mathbf{f}_{BB}^H\mathbf{F}_{RF}^H\mathbf{A}_{te}^H\mathbf{A}_{te}\mathbf{F}_{RF}\mathbf{f}_{BB}}{\lambda_2\mathrm{tr}[\mathbf{A}_{te}\mathbf{F}_{RF}\mathrm{diag}\{\mathbf{f}_{BB}\mathbf{f}_{BB}^H\}\mathbf{F}_{RF}^H\mathbf{A}_{te}^H]+\gamma_e}\nonumber\\
&=\frac{\lambda_1\mathbf{f}_{BB}^H\tilde{\mathbf{A}}_{te}^H\tilde{\mathbf{A}}_{te}\mathbf{f}_{BB}}{\lambda_2\mathrm{tr}[\tilde{\mathbf{A}}_{te}(\mathbf{f}_{BB}\mathbf{f}_{BB}^H)\odot\mathbf{I}_K\tilde{\mathbf{A}}_{te}^H]+\gamma_e}\nonumber\\
&=\frac{\lambda_1\mathbf{f}_{BB}^H\tilde{\mathbf{A}}_{te}^H\tilde{\mathbf{A}}_{te}\mathbf{f}_{BB}}{\lambda_2\mathrm{tr}[(\mathbf{f}_{BB}\mathbf{f}_{BB}^H)\{\mathbf{I}_K\odot(\tilde{\mathbf{A}}_{te}^H\tilde{\mathbf{A}}_{te})\}]+\gamma_e}\nonumber\\
&=\frac{\mathbf{f}_{BB}^H(\lambda_1\tilde{\mathbf{A}}_{te}^H\tilde{\mathbf{A}}_{te})\mathbf{f}_{BB}}{\mathbf{f}_{BB}^H(\lambda_2[\mathbf{I}_K\odot\tilde{\mathbf{A}}_{te}^H\tilde{\mathbf{A}}_{te}]+\gamma_e\mathbf{I}_K)\mathbf{f}_{BB}}
\end{align}
where
\begin{align}
\lambda_1=\beta P(1-\eta)^2N_t/L,
\end{align}
\begin{align}
\lambda_2=\eta(1-\eta)\beta P N_t/L,
\end{align}
\begin{align}
&\gamma_e=(1-\beta)P(1-\eta)^2\frac{N_t}{L}\mathrm{tr}[\mathbf{A}_{te}\mathbf{F}_{RF}\mathbf{T}_{BB}\mathbf{T}_{BB}^H\mathbf{F}_{RF}^H\mathbf{A}_{te}^H]\!+\!\nonumber\\
&\!\frac{N_t}{L}\!\mathrm{tr}[\mathbf{A}_{te}\mathbf{F}_{RF}\eta(1\!-\!\eta)\mathrm{diag}\!\{\!(1\!-\!\beta)P\mathbf{T}_{BB}\mathbf{T}_{BB}^H\!\}\!\mathbf{F}_{RF}^H\mathbf{A}_{te}^H]\!+\!\sigma^2, \end{align}
and $\tilde{\mathbf{A}}_{te}=\mathbf{A}_{te}\mathbf{F}_{RF}$. Using the above expression of  $\mathbb{E}\{\mathrm{SINR}_{e}\}$ , we directly have the average achievable rate at Eve as follows:
\begin{align}\label{Re-f-BB}
\log_2\left(1+\mathbb{E}[\mathrm{SINR}_e]\right)\approx\log_2\left(\frac{\mathbf{f}_{BB}^H\mathbf{Q}_e\mathbf{f}_{BB}}{\mathbf{f}_{BB}^H\mathbf{P}_e\mathbf{f}_{BB}}
\right)
\end{align}
where
\begin{align}
\mathbf{Q}_{e}=\lambda_2[\mathbf{I}_K\odot\tilde{\mathbf{A}}_{te}^H\tilde{\mathbf{A}}_{te}]+\gamma_e\mathbf{I}_K+\lambda_1\tilde{\mathbf{A}}_{te}^H\tilde{\mathbf{A}}_{te} \end{align}
and
\begin{align}
\mathbf{P}_{e}=\lambda_2[\mathbf{I}_K\odot\tilde{\mathbf{A}}_{te}^H\tilde{\mathbf{A}}_{te}]+\gamma_e\mathbf{I}_K.
\end{align}
Combining (\ref{Rb-f-BB})and (\ref{Re-f-BB}) yields the expression of $\tilde{R}_s$  as
\begin{align}\label{Rs_fBB}
&\tilde{R}_s=R_b-\log_2\left(1+\mathbb{E}[\mathrm{SINR}_e]\right)\nonumber\\
&\approx\log_2\left(\frac{\mathbf{f}_{BB}^H\mathbf{Q}_b\mathbf{f}_{BB}}{\mathbf{f}_{BB}^H\mathbf{P}_b\mathbf{f}_{BB}}\cdot\frac{\mathbf{f}_{BB}^H\mathbf{P}_{e}\mathbf{f}_{BB}}{\mathbf{f}_{BB}^H\mathbf{Q}_{e}\mathbf{f}_{BB}}\right)
\end{align}

Observing the form of (\ref{Rs_fBB}) is the product of fractional quadratic functions, one efficient way to address this problem is GPI algorithm \cite{Lee}. Therefore, we can obtain the solution to $\mathbf{f}_{BB}$ by resorting to GPI algorithm.

\subsection{Design of ANPM}

In this subsection, we will optimize the digital ANPM by fixing $\mathbf{F}_{RF}$ and $\mathbf{f}_{BB}$. Under this condition, we can rewrite the $R_b$ as a function of $\mathbf{T}_{BB}$ as follows:
\begin{align}\label{Rb-TBB}
&R_b(\mathbf{T}_{BB})\nonumber\\
&\!=\!\mathrm{log}_2\left(1\!+\!\frac{\kappa_b}{\alpha_1\|\tilde{\mathbf{h}}_b\mathbf{T}_{BB}\|^2\!+\!\alpha_2\tilde{\mathbf{h}}_b\mathrm{diag}\{\mathbf{T}_{BB}\mathbf{T}_{BB}^H\}\tilde{\mathbf{h}}_b^H\!+\!\omega_b}\right)\nonumber\\
&\!=\!\mathrm{log}_2\!\left(\!1\!+\!\frac{\kappa_b}{\alpha_1\|\tilde{\mathbf{h}}_b\!\mathbf{T}_{BB}\!\|^2\!+\!\alpha_2\mathrm{tr}\!\{\!\mathbf{T}_{BB}^H[\mathbf{I}\!\odot\!(\tilde{\mathbf{h}}_b^H\tilde{\mathbf{h}}_b)]\mathbf{T}_{BB}\!\}\!\!+\!\omega_b}\!\right)\!\nonumber\\
&\!=\!\mathrm{log}_2\left(1+\frac{\kappa_b}{\alpha_1\|\tilde{\mathbf{h}}_b\mathbf{T}_{BB}\|^2+\alpha_2\mathrm{tr}\{\mathbf{T}_{BB}^H\mathbf{C}_b\mathbf{T}_{BB}\}+\omega_b}\right)\nonumber\\
&\!=\!\mathrm{log}_2\left(\frac{\mathrm{Tr}\{\mathbf{T}_{BB}^H\mathbf{E}_b\mathbf{T}_{BB}\}}{\mathrm{Tr}\{\mathbf{T}_{BB}^H\mathbf{F}_b\mathbf{T}_{BB}\}}\right)\nonumber\\
&\!=\!\mathrm{log}_2\left(\frac{\mathbf{w}^H\left(\mathbf{I}\otimes\mathbf{E}_b\right)\mathbf{w}}{\mathbf{w}^H\left(\mathbf{I}\otimes\mathbf{F}_b\right)\mathbf{w}}\right)
\end{align}
where
\begin{align}
  \mathbf{w}=\mathrm{vec}\left(\mathbf{T}_{BB}\right),
  \end{align}
 \begin{align}
 \alpha_1=(1-\beta)P(1-\eta)^2,
 \end{align}
  \begin{align}
  \alpha_2=\eta(1-\eta)(1-\beta)P,
  \end{align}

\begin{align}
&\kappa_b=\beta P(1-\eta)^2\mathbf{h}_b\mathbf{F}_{RF}\mathbf{f}_{BB}\mathbf{f}_{BB}^H\mathbf{F}_{RF}^H\mathbf{h}_b^H,
\end{align}
\begin{align} &\omega_b=\eta(1-\eta)\beta P\tilde{\mathbf{h}}_b\mathrm{diag}[\mathbf{f}_{BB}\mathbf{f}_{BB}^H]\tilde{\mathbf{h}}_b^H+\sigma^2,
\end{align}

\begin{align}
\mathbf{C}_b=\mathbf{I}\odot(\tilde{\mathbf{h}}_b^H\tilde{\mathbf{h}}_b),
 \end{align}
\begin{align}
\mathbf{E}_b=\alpha_1\tilde{\mathbf{h}}_b^H\tilde{\mathbf{h}}_b+\alpha_2\mathbf{C}_b+\omega_b\mathbf{I}_K+\kappa_b\mathbf{I}_K,
 \end{align}
 and
\begin{align}
\mathbf{F}_b=\alpha_1\tilde{\mathbf{h}}_b^H\tilde{\mathbf{h}}_b+\alpha_2\mathbf{C}_b+\omega_b\mathbf{I}_K.
\end{align}

To obtain the $\tilde{R}_s$,  $\mathbb{E}\{\mathrm{SINR}_e\}$ is rewritten as a function of $\mathbf{T}_{BB}$ as follows:
\begin{align}
&\mathop{\mathbb{E}[\mathrm{SINR}_{e}]}\limits_{\mathbf{T}_{BB}}\nonumber\\
&\!\approx\!\frac{\kappa_e}{\alpha_3\mathrm{tr}[\!\mathbf{T}_{BB}^H\!\tilde{\mathbf{A}}_{te}^H\tilde{\mathbf{A}}_{te}\!\mathbf{T}_{BB}\!]\!+\!\!\alpha_4\!\mathrm{tr}[\mathbf{T}_{BB}^H[\mathbf{I}_K\!\odot\!(\tilde{\mathbf{A}}_{te}^H\tilde{\mathbf{A}}_{te})]\mathbf{T}_{BB}]\!+\!\omega_e}\nonumber\\
&=\frac{\kappa_e}{\mathrm{tr}[\mathbf{T}_{BB}^H\{\alpha_3\tilde{\mathbf{A}}_{te}^H\tilde{\mathbf{A}}_{te}+\alpha_4\mathbf{C}_e+\omega_e\mathbf{I}_K\}\mathbf{T}_{BB}]}\nonumber\\
&=\frac{\kappa_e}{\mathrm{tr}[\mathbf{T}_{BB}^H\mathbf{F}_e\mathbf{T}_{BB}]}
\end{align}
where
\begin{align}
\alpha_3=(1-\beta)P(1-\eta)^2\frac{N_t}{L}, \alpha_4=\eta(1-\eta)(1-\beta)P\frac{N_t}{L},
\end{align}
\begin{align}
\kappa_e=\beta P(1-\eta)^2\frac{N_t}{L}\mathrm{tr}[\tilde{\mathbf{A}}_{te}\mathbf{f}_{BB}\mathbf{f}_{BB}^H\tilde{\mathbf{A}}_{te}], \end{align}
and
\begin{align}
\omega_e=\eta(1-\eta)\beta P\frac{N_t}{L}\mathrm{tr}[\tilde{\mathbf{A}}_{te}\mathrm{diag}(\mathbf{f}_{BB}\mathbf{f}_{BB}^H)\tilde{\mathbf{A}}_{te}^H]+\sigma^2.
\end{align}

Therefore, we can obtain
\begin{align}\label{Re-TBB}
&\log_2\left(1+\mathbb{E}[\mathrm{SINR}]\right)\approx\log_2\left(\frac{\mathrm{tr}[\mathbf{T}_{BB}^H(\mathbf{F}_e+\kappa_e\mathbf{I}_K)\mathbf{T}_{BB}]}{\mathrm{tr}[\mathbf{T}_{BB}^H\mathbf{F}_e\mathbf{T}_{BB}]}\right)\nonumber\\
&=\log_2\left(\frac{\mathrm{tr}[\mathbf{T}_{BB}^H\mathbf{E}_e\mathbf{T}_{BB}]}{\mathrm{tr}[\mathbf{T}_{BB}^H\mathbf{F}_e\mathbf{T}_{BB}]}\right)\nonumber\\
&=\log_2\left(\frac{\mathbf{w}^H\left(\mathbf{I}\otimes\mathbf{E}_e\right)\mathbf{w}}{\mathbf{w}^H\left(\mathbf{I}\otimes\mathbf{F}_e\right)\mathbf{w}}\right)
\end{align}
with
\begin{align}
\mathbf{C}_e={\mathbf{I}_K\odot(\tilde{\mathbf{A}}_{te}^H\tilde{\mathbf{A}}_{te})},
\end{align}
\begin{align}
\mathbf{E}_e=\alpha_3\tilde{\mathbf{A}}_{te}^H\tilde{\mathbf{A}}_{te}+\alpha_4{\mathbf{I}_K\odot(\tilde{\mathbf{A}}_{te}^H\tilde{\mathbf{A}}_{te})}+\omega_e\mathbf{I}_K+\kappa_e\mathbf{I}_K, \end{align}
and
\begin{align}
\mathbf{F}_e=\alpha_3\tilde{\mathbf{A}}_{te}^H\tilde{\mathbf{A}}_{te}+\alpha_4{\mathbf{I}_K\odot(\tilde{\mathbf{A}}_{te}^H\tilde{\mathbf{A}}_{te})}+\omega_e\mathbf{I}_K.
\end{align}
Making use of (\ref{Rb-TBB}) and (\ref{Re-TBB}), the $\tilde{R}_s$ can be reformulated as a function of $\mathbf{w}$ with $\mathbf{F}_{RF}$ and $\mathbf{f}_{BB}$ fixed as follows:
\begin{align}\label{Rs_TBB}
&\tilde{R}_s=R_b-\log_2\left(1+\mathbb{E}[\mathrm{SINR}]\right)\nonumber\\
&\approx\log_2\left(\frac{\mathbf{w}^H\left(\mathbf{I}\otimes\mathbf{E}_b\right)\mathbf{w}}{\mathbf{w}^H\left(\mathbf{I}\otimes\mathbf{F}_b\right)\mathbf{w}}\cdot\frac{\mathbf{w}^H\left(\mathbf{I}\otimes\mathbf{F}_e\right)\mathbf{w}}{\mathbf{w}^H\left(\mathbf{I}\otimes\mathbf{E}_e\right)\mathbf{w}}\right)
\end{align}
where $\mathbf{w}=\mathrm{vec}\left(\mathbf{T}_{BB}\right)$.

\begin{algorithm}[htb]
\caption{Iterative method for digital precoders}
\textbf{Input}:\\
(1) Fixed $\mathbf{F}_{RF}$, initialize $\mathbf{f}_{BB}^{(0)}$ and $\mathbf{T}_{BB}^{(0)}$ and set $k=0$, threshold value $\epsilon$;\\
(2) Compute $R_s^{(0)}$ using $\mathbf{f}_{BB}^{(0)}$, $\mathbf{T}_{BB}^{(0)}$ and $\mathbf{F}_{RF}$, search the optimal $\beta^{0}$ by 1-D search;\\
~\textbf{Repeat}
\begin{algorithmic}[1] 
\STATE Solve $\mathbf{T}_{BB}^{(k+1)}$ using $\mathbf{f}_{BB}^{(k)}$ and $\mathbf{F}_{RF}$ through (\ref{Rs_TBB});\\
\STATE Solve $\mathbf{f}_{BB}^{(k+1)}$ using $\mathbf{T}_{BB}^{(k)}$ and $\mathbf{F}_{RF}$ through (\ref{Rs_fBB});\\
\STATE Compute $R_s^{(k+1)}$ using $\mathbf{T}_{BB}^{(k+1)}$, $\mathbf{f}_{BB}^{(k+1)}$ and $\mathbf{F}_{RF}$, search the optimal $\beta^{k+1}$ by 1-D search;\\
\STATE $k=k+1$;\\
\end{algorithmic}
~\textbf{Until} $R_s^{(k+1)}-R_s^{(k)}\leq\epsilon$;\\
\textbf{Output}: $\mathbf{f}_{BB}^{(k)}$ and $\mathbf{T}_{BB}^{(k)}$;
\end{algorithm}

\begin{algorithm}[htb]
\caption{Proposed TLAIS algorithm}
\textbf{Input}:\\
(1) Initialize $\mathbf{F}_{RF}^{(0)}$, $\mathbf{f}_{BB}^{(0)}$ and $\mathbf{T}_{BB}^{(0)}$ according to (\ref{Frf_initial}), (\ref{Fbb_initial}) and (\ref{TBB_initial}), respectively and set $p=0$, threshold value $\epsilon$;\\
(2) For $\mathbf{F}_{RF}^{(0)}$, update $\mathbf{f}_{BB}^{(0)}$ and $\mathbf{T}_{BB}^{(0)}$ with Algorithm 2;\\
~\textbf{Repeat}
\begin{algorithmic}[1] 
\STATE For $\mathbf{f}_{BB}^{(p)}$ and $\mathbf{T}_{BB}^{(p)}$, using Algorithm 1 to obtain $\mathbf{F}_{RF}^{(p+1)}$;\\
\STATE For $\mathbf{F}_{RF}^{(p+1)}$, using Algorithm 2 to obtain updated $\mathbf{f}_{BB}^{(p+1)}$ and $\mathbf{T}_{BB}^{(p+1)}$;\\
\STATE Compute $R_s^{(p+1)}$ using $\mathbf{T}_{BB}^{(p+1)}$, $\mathbf{f}_{BB}^{(p+1)}$ and $\mathbf{F}_{RF}^{(p+1)}$;\\
\STATE $p=p+1$;\\
\end{algorithmic}
~\textbf{Until} $R_s^{(p+1)}-R_s^{(p)}\leq\epsilon$;\\
\textbf{Output}: $R_s^{(p)}$.
\end{algorithm}
Similarly, $\mathbf{w}$ can be solved with GPI, and the corresponding ANPM  can be obtained by reverse operation of vector-to-matrix operator.

Until now, we have completed the design of secure hybrid precoders. Our iterative idea can be described as follows: for given AP matrix $\mathbf{F}_{RF}$,  the near-optimal DP vector $\mathbf{f}_{BB}$ and ANPM $\mathbf{T}_{BB}$ are iteratively computed; Then given  the DP vector $\mathbf{f}_{BB}$ and ANPM $\mathbf{T}_{BB}$, we renew the AP matrix $\mathbf{F}_{RF}$ by GA method.  The alternative iterations among $\mathbf{F}_{RF}$, $\mathbf{f}_{BB}$, and $\mathbf{T}_{BB}$ is repeated until a stop criterion satisfies. Here, stop criterion is chosen as follows: $R_s^{(p+1)}-R_s^{(p)}\leq\epsilon$ with $p$ being the iteration index. The proposed method is summarized in Algorithm 3.
\subsection{Initialization of TLAIS}
In the following, we will show how to set the initial values of  the  proposed TLAIS. In other words, we need to make an initialization for three matrices AP, DP, and ANPM, respectively. Since the optimization problem in  (\ref{Rs_low}) is non-convex and intractable to be directly handled. Thus, to simplify it, we first consider the scenario without AN, i.e.  $\beta=1$,  as in \cite{Y.R}. In this case, the achievable rate of Bob can be expressed as
\begin{align}\label{analog_Rb}
\hspace{-2mm}
R_b&\!=\!\log_2\!\left(\!1\!+\!\frac{P(1-\eta)^2\|\mathbf{h}_b\mathbf{F}_{RF}\mathbf{f}_{BB}\|^2}{\eta(1-\eta) P\mathbf{h}_b\mathbf{F}_{RF}\mathrm{diag}\{\mathbf{f}_{BB}\mathbf{f}_{BB}^H\}\mathbf{F}_{RF}^H\mathbf{h}_b^H+\sigma_b^2}\!\right)\!\nonumber\\
&\!=\!\log_2\!\left(\!1+\frac{\mu_1|\mathbf{h}_b\mathbf{F}_{RF}\mathbf{f}_{BB}\|^2}{\mu_2\mathbf{h}_b\mathbf{F}_{RF}[(\mathbf{f}_{BB}\mathbf{f}_{BB}^H)\odot\mathbf{I}_K]\mathbf{F}_{RF}^H\mathbf{h}_b^H+\sigma_b^2}\right)\nonumber\\
&\!=\!\log_2\!\left(\!1+\frac{\mu_1\mathbf{f}_{BB}^H\mathbf{F}_{RF}^H\mathbf{h}_b^H\mathbf{h}_b\mathbf{F}_{RF}\mathbf{f}_{BB}}{\mu_2\mathbf{f}_{BB}^H[\mathbf{I}_K\odot\left(\mathbf{F}_{RF}^H\mathbf{h}_b^H\mathbf{h}_b\mathbf{F}_{RF}\right)]\mathbf{f}_{BB}+\sigma_b^2}\right)
\end{align}
where $\mu_1=P(1-\eta)^2$ and $\mu_2=\eta(1-\eta)P$.

If a fully digital architecture is adopted, i.e., $\mathbf{F}_{RF}=\mathbf{I}_{Nt}$, then (\ref{analog_Rb}) can be simplified as
\begin{align}
R_b&=\log_2\!\left(\!1+\frac{\mu_1\mathbf{f}_{FD}^H\mathbf{h}_b^H\mathbf{h}_b\mathbf{f}_{FD}}{\mu_2\mathbf{f}_{FD}^H[\mathbf{I}_{N_t}\odot\left(\mathbf{h}_b^H\mathbf{h}_b\right)]\mathbf{f}_{FD}+\sigma_b^2}\right)\nonumber\\
&=\log_2\left(\frac{\mathbf{f}_{FD}^H\mathbf{X}_b\mathbf{f}_{FD}}{\mathbf{f}_{FD}^H\mathbf{Y}_b\mathbf{f}_{FD}}\right)
\end{align}
where
\begin{align}
\mathbf{X}_b=\mu_2[\mathbf{I}_{Nt}\odot(\mathbf{h}_b^H\mathbf{h}_b)]+\sigma^2\mathbf{I}_{N_t}+\mu_1\mathbf{h}_b^H\mathbf{h}_b,
\end{align}
and
\begin{align}
\mathbf{Y}_b=\mu_2[\mathbf{I}_{Nt}\odot(\mathbf{h}_b^H\mathbf{h}_b)]+\sigma^2\mathbf{I}_{N_t}.
\end{align}

Similarly, the $\log_2\left(1+\mathbb{E}[\mathrm{SINR_e}]\right)$ at Eve can be rewritten as
\begin{align}\label{RB-Initial}
&\log_2\left(1+\mathbb{E}[\mathrm{SINR}_e]\right)\nonumber\\
&\overset{(f)}\approx\log_2\left(1+\frac{\mu_3\mathbf{f}_{BB}^H\mathbf{F}_{RF}^H\mathbf{A}_{te}^H\mathbf{A}_{te}\mathbf{F}_{RF}\mathbf{f}_{BB}}{\mu_4\mathbf{f}_{BB}^H[\mathbf{I}_K\odot(\mathbf{F}_{RF}^H\mathbf{A}_{te}^H\mathbf{A}_{te}\mathbf{F}_{RF})]\mathbf{f}_{BB}+\sigma^2}\right)\nonumber\\
&\overset{(g)}=\log_2\left(1+\frac{\mu_3\mathbf{f}_{FD}^H\mathbf{A}_{te}^H\mathbf{A}_{te}\mathbf{f}_{FD}}{\mu_4\mathbf{F}_{FD}^H[\mathbf{I}_K\odot(\mathbf{A}_{te}^H\mathbf{A}_{te})]\mathbf{f}_{FD}+\sigma^2}\right)\nonumber\\
&=\log_2\left(\frac{\mathbf{f}_{FD}\mathbf{X}_e\mathbf{f}_{FD}}{\mathbf{f}_{FD}^H\mathbf{Y}_e\mathbf{f}_{FD}}\right)
\end{align}
where $\mu_3=P(1-\eta)^2\frac{N_t}{L}$, $\mu_4=\eta(1-\eta)P\frac{N_t}{L}$. The approximation $(f)$ in (\ref{RB-Initial}) holds because $\beta=1$ and $(g)$ is true due to $\mathbf{F}_{RF}=\mathbf{I}_{N_t}$. What's more,
\begin{align}
\mathbf{X}_e=\mu_4[\mathbf{I}_{N_t}\odot(\mathbf{A}_{te}^H\mathbf{A}_{te})]+\sigma^2\mathbf{I}_{N_t}+\mu_3\mathbf{A}_{te}^H\mathbf{A}_{te},
\end{align}
and
\begin{align}
\mathbf{Y}_e=\mu_4[\mathbf{I}_{N_t}\odot(\mathbf{A}_{te}^H\mathbf{A}_{te})]+\sigma^2\mathbf{I}_{N_t}
\end{align}
Therefore, the approximate expression $\tilde{R}_s$ of SR  can be expressed as
\begin{align}\label{Rs_Frf}
&\tilde{R}_s=R_b-\log_2\left(1+\mathbb{E}[\mathrm{SINR}]\right)\nonumber\\
&\approx\log_2\left(\frac{\mathbf{f}_{FD}^H\mathbf{X}_b\mathbf{f}_{FD}}{\mathbf{f}_{FD}^H\mathbf{Y}_b\mathbf{f}_{FD}}\cdot\frac{\mathbf{f}_{FD}^H\mathbf{Y}_b\mathbf{f}_{FD}}{\mathbf{f}_{FD}^H\mathbf{X}_b\mathbf{f}_{FD}}\right).
\end{align}

Therefore, $\mathbf{f}_{FD}$ can be solved with GPI algorithm. After obtaining $\mathbf{f}_{FD}$, we can obtain analog beamforming $\mathbf{F}_{RF}$ as
\begin{align}\label{Frf_initial}
\mathbf{f}_{RF,l}=\frac{1}{\sqrt{T}}\mathrm{exp}\left\{jQ_{bps}[\mathbf{f}_{FD,(l-1)M:lM}]\right\},l=1,2,\cdots,K.
\end{align}
where $Q_{bps}$ denotes the operation of quantizing each phase of the $\mathbf{f}_{FD}$ to its nearest value in discrete phase set $\mathcal{F}_{RF}$.

Observing (\ref{analog_Rb}), we can assign the initial value of $\mathbf{f}_{BB}$ by rewriting (\ref{analog_Rb}) as
\begin{align}
R_b&=\log_2\left(1+\frac{\mathbf{f}_{BB}^H[\mu_1\mathbf{F}_{RF}^H\mathbf{h}_b^H\mathbf{h}_b\mathbf{F}_{RF}]\mathbf{f}_{BB}}{\mathbf{f}_{BB}^H\{\mu_2[\mathbf{I}_K\odot(\mathbf{F}_{RF}^H\mathbf{h}_b^H\mathbf{h}_b\mathbf{F}_{RF})]+\sigma^2\}\mathbf{f}_{BB}}\right)\nonumber\\
&=\log_2\left(\frac{\mathbf{f}_{BB}^H\tilde{\mathbf{X}}_b\mathbf{f}_{BB}}{\mathbf{f}_{BB}^H\tilde{\mathbf{Y}}_b\mathbf{f}_{BB}}\right).
\end{align}
Similarly, we have the approximate rate at Eve as
\begin{align}
&\log_2\left(1+\mathbb{E}[\mathrm{SINR}_e]\right)\nonumber\\
&\approx\log_2\left(1+\frac{\mu_3\mathbf{f}_{BB}^H\mathbf{F}_{RF}^H\mathbf{A}_{te}^H\mathbf{A}_{te}\mathbf{F}_{RF}\mathbf{f}_{BB}}{\mu_4\mathbf{f}_{BB}^H[\mathbf{I}_K\odot(\mathbf{F}_{RF}^H\mathbf{A}_{te}^H\mathbf{A}_{te}\mathbf{F}_{RF})]\mathbf{f}_{BB}+\sigma^2}\right)\nonumber\\
&=\log_2\left(\frac{\mathbf{f}_{BB}^H\tilde{\mathbf{X}}_e\mathbf{f}_{BB}^H}{\mathbf{f}_{BB}^H\tilde{\mathbf{Y}}_e\mathbf{f}_{BB}}\right)
\end{align}
with
\begin{align}
\tilde{\mathbf{X}}_b\!=\!\mu_2[\mathbf{I}_K\!\odot\!(\mathbf{F}_{RF}^H\mathbf{h}_b^H\mathbf{h}_b\mathbf{F}_{RF})]\!+\!\sigma^2\mathbf{I}_K\!+\!\mu_1\mathbf{F}_{RF}^H\mathbf{h}_b^H\mathbf{h}_b\mathbf{F}_{RF},
\end{align}
\begin{align}
\tilde{\mathbf{Y}}_b=\mu_2[\mathbf{I}_K\odot(\mathbf{F}_{RF}^H\mathbf{h}_b^H\mathbf{h}_b\mathbf{F}_{RF})]+\sigma^2\mathbf{I}_K,
\end{align}
\begin{align}
\tilde{\!\mathbf{X}}_e\!\!=\!\mu_4[\mathbf{I}_K\!\odot\!(\mathbf{F}_{RF}^H\mathbf{A}_{te}^H\mathbf{A}_{te}\!\mathbf{F}_{RF}\!)]\!+\!\sigma^2\mathbf{I}_K\!+\!\mu_3\mathbf{F}_{RF}^H\mathbf{A}_{te}^H\mathbf{A}_{te}\!\mathbf{F}_{RF}\!,
\end{align}
and
\begin{align}
\tilde{\mathbf{X}}_e=\mu_4[\mathbf{I}_K\odot(\mathbf{F}_{RF}^H\mathbf{A}_{te}^H\mathbf{A}_{te}\mathbf{F}_{RF})]+\sigma^2\mathbf{I}_K.
\end{align}
Then the $\tilde{R}_s$ can be rewritten as
\begin{align}\label{Fbb_initial}
&\tilde{R}_s=R_b-\log_2\left(1+\mathbb{E}[\mathrm{SINR}]\right)\nonumber\\
&\approx\log_2\left(\frac{\mathbf{f}_{BB}^H\tilde{\mathbf{X}}_b\mathbf{f}_{BB}}{\mathbf{f}_{BB}^H\tilde{\mathbf{Y}}_b\mathbf{f}_{BB}}\cdot\frac{\mathbf{f}_{BB}^H\tilde{\mathbf{Y}}_e\mathbf{f}_{BB}}{\mathbf{f}_{BB}^H\tilde{\mathbf{X}}_e\mathbf{f}_{BB}}\right)
\end{align}
Therefore, the initial value of $\mathbf{f}_{BB}$ can be obtained by maximizing (\ref{Fbb_initial}) via GPI scheme.

To set a good initial value of ANPM, it's reasonable to assign the initial value of $\mathbf{T}_{BB}$ by the concept of null-space projection, which forces the $\mathbf{T}_{BB}$ in the null space of channel of desired user, thus eliminating the interference of AN to the intended user. The expression of the initial value of $\mathbf{T}_{BB}$ can be given as follows:
\begin{align}\label{TBB_initial}
\mathbf{T}_{BB}=\frac{\mathbf{I}_{K}-\tilde{\mathbf{h}}_{b}^H\left(\tilde{\mathbf{h}}_b\tilde{\mathbf{h}}_b^H\right)^{-1}\tilde{\mathbf{h}}_b}{\|\mathbf{I}_{K}-\tilde{\mathbf{h}}_{b}^H\left(\tilde{\mathbf{h}}_b\tilde{\mathbf{h}}_b^H\right)^{-1}\tilde{\mathbf{h}}_b\|_F}
\end{align}
where $\tilde{\mathbf{h}}_b=\mathbf{h}_b\mathbf{F}_{RF}$ stands for the equivalent channel of intended user.

Since the beamforming matrices $\mathbf{f}_{BB}$ and $\mathbf{T}_{BB}$ are closely related to the power allocation factor $\beta$ in accordance with (\ref{Rs_fBB}) and (\ref{Rs_TBB}), the reasonable way to find the optimal PA is achieved by the conventional 1-D linear search.

\section{Numerical Results}
In this section, we will present the numerical simulations to evaluate the SR performance of our proposed method and compare it with existing methods like AN-aided method in \cite{Y.R}. Here it is particularly pointed out that since it's infeasible to directly apply the algorithm without taking low-resolution DACs into account in \cite{Y.R} to our paper. Then,  the idea in \cite{Y.R} is modified to suit our model and at the same time choose it as our initial value of our proposed precoders. The second algorithm is the maximum ratio transmission (MRT) algorithm. To guarantee the fair of performance comparison,  the same AN power is introduced to MRT. What's more, to reveal the importance of AN, we simultaneously make a comparison of our proposed method without AN-aided together with MRT algorithm.
The parameters are chosen as follows. In our simulation experiment, the transmitter at Alice is equipped with 32 transmit antennas with 4 RF chains, and mmWave channel model is also adopted in our paper with $L=12$ and azimuth angles being a uniform distribution over $[0, 2\pi]$.

\begin{figure}[h]
\centering
\includegraphics[width=8cm]{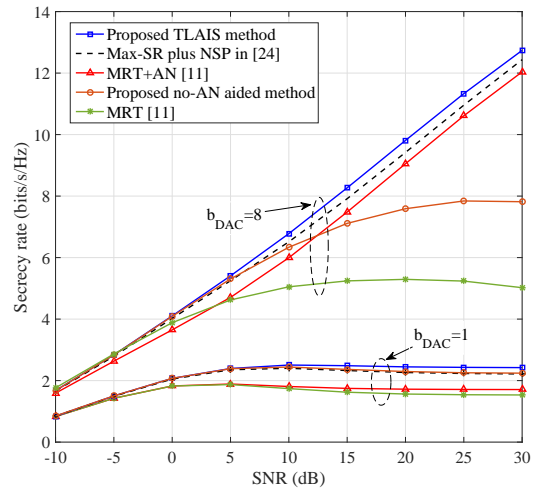}\\
\caption{SR versus SNR with $b_{DAC}=8$ bits.}
\label{sr}
\end{figure}
Fig.~\ref{sr} demonstrates the SR performance versus SNR with number $b_{DAC}$ of quantization bits of ADC being 8. As can be seen from this figure, the proposed TLAIS method performs better than AN-aided method in \cite{Y.R} and MRT plus AN method in terms of SR performance. From this figure, we find an important fact that AN has an important impact on SR performance. Taking the proposed TLAIS as an example, the proposed TLAIS performs much better than the corresponding one without AN. More important, with the help of AN, the SR performance of all three methods including our proposed method gradually grows  with increase in SNR.  Conversely, without the aid of AN, all methods exist the effect of SR performance ceil as SNR increases. The main reason that our proposed method is better than MRT can be explained as follows: unlike our method, the major goal of MRT algorithm is to maximize the receive SNR at Bob and don't concern the receive SNR at Eve. However, our method is to maximize SR. Thus, our proposed method can achieve a better SR performance.

\begin{figure}[h]
\centering
\includegraphics[width=8cm]{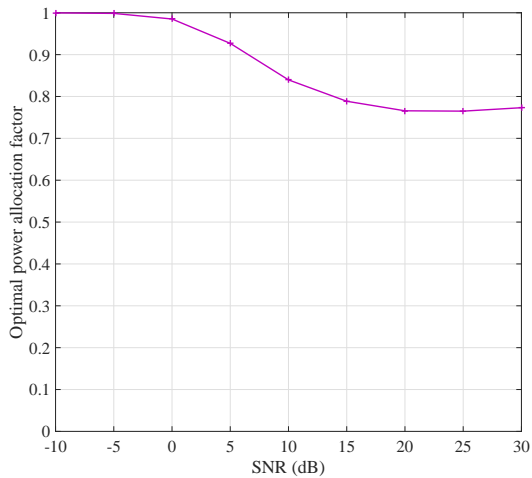}\\
\caption{Optimal PA factor versus SNR.}
\label{PA}
\end{figure}
Fig.~\ref{PA} shows the optimal PA factor $\beta$ versus SNR with $b_{DAC}=8$, where $\beta$ is to maximize SR by adjusting PA between confidential message and AN. From  Fig.~\ref{PA}, it is seen that, in the low SNR region, the optimal PA factor always remains 1, which implies that in this scenario all power should be allocated to transmit confidential message to guarantee the communication quality, and AN has a trivial impact on SR performance. Furthermore, as  SNR increases, the optimal PA factor declines gradually. This  reveals a fact that more power should be allocated to AN to interfere with eavesdropper, and improves the security of communication. Fig.~\ref{PA} verifies the effectiveness of AN-aided method especially in the medium and high SNR regions.

\begin{figure}[h]
\centering
\includegraphics[width=8cm]{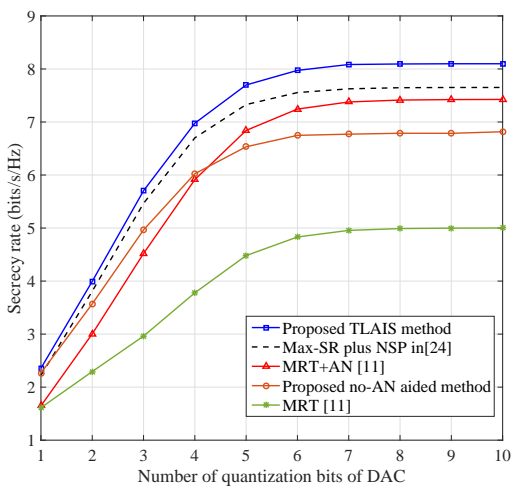}\\
\caption{SR versus number of quantization bits of DAC with SNR=15dB.}
\label{SR_QDAC}
\end{figure}
Fig.~\ref{SR_QDAC} illustrates  the SR versus number $b_{DAC}$  of quantization bits of DAC with $\mathrm{SNR=15dB}$. For the small value of $b_{DAC}$, the SR is small and  its performance loss is significant because the distortion of low-resolution DACs is extremely severe. As  $b_{DAC}$  increases, the SR performance improves dramatically, especially when $b_{DAC}$ ranges from $1$ to $6$ bits. It's noted that the SR performance of five methods all tend to stable when $b_{DAC}$ reach up to 6. In such a situation,  the performance loss is negligible. What's more, given a fixed value of $b_{DAC}$, our proposed method achieves a higher SR performance than other algorithms.

\begin{figure}[h]
\centering
\includegraphics[width=8cm]{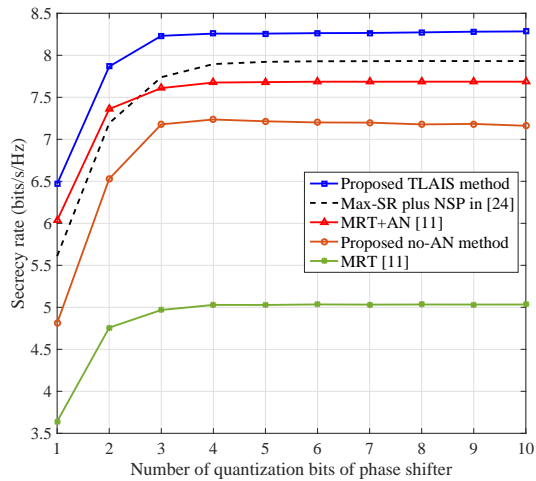}\\
\caption{SR versus number of quantization bits of phase shift with SNR=15dB.}
\label{SR_QPS}
\end{figure}
Fig.~\ref{SR_QPS} plots the SR versus number $b_{PS}$ of quantization bits of phase shifters with $\mathrm{SNR=15dB}$ and $b_{DAC}=8$.   The proposed TLAIS performs much better than MRT and MRT with AN in terms of SR for all cases. Additionally, we also find the fact that their SR tend to a flat rate ceil  with  increase in number of quantization bits of phase shifts. From Fig.~\ref{SR_QPS}, it follows that the SR loss due to the effect of quantization error from phase shifters becomes trivial when $b_{PS}\geq4$ .

\begin{figure}[h]
\centering
\includegraphics[width=8cm]{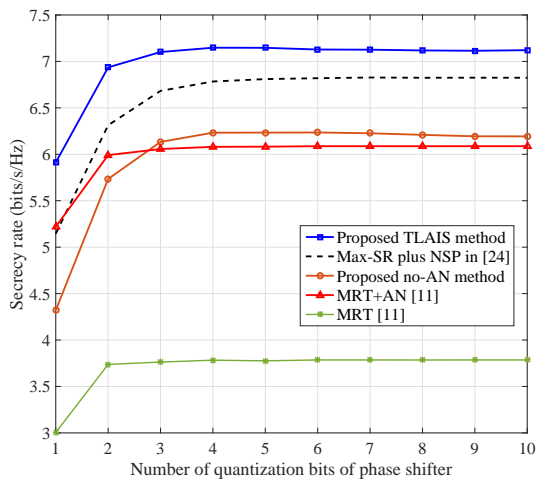}\\
\caption{SR versus number of quantization bits of phase shifter with SNR=15dB and $b_{DAC}=4$.}
\label{SR_QPS_dac}
\end{figure}

Fig.~\ref{SR_QPS_dac} plots the SR versus number $b_{PS}$ of quantization bits of phase shifters with $\mathrm{SNR=15dB}$ and $b_{DAC}=4$. It is noted that Fig.~\ref{SR_QPS_dac} shows the similar performance trend with Fig.~\ref{SR_QPS}. In other words, it follows that our proposed no-AN aided method is still better than MRT plus AN method in medium and high values of $b_{PS}$. This result is consistent  with that Fig.~\ref{SR_QDAC} at  $b_{DAC}=4$.

\section{Conclusion}
In this paper,  a TLAIS was proposed  for AN-aided hybrid precoding by taking the low-resolution DACs and finite-quantized phase shifters into account. First, we developed a quantized DAC model for AN aided hybrid precoding based on AQN model including the effect of quantization noise. Then,  an approximate expression of SR with partial channel knowledge of eavesdropper was also derived. Based on the derived approximate expression,a TLAIS was presented to optimize the design of DP vector, ANPM, and AP matrix by resorting to the principle of GPI. Considering the unit modulus constraints of AP, we adopted a GA algorithm to compute the AP matrix. Furthermore, to reduce the computation complexity, only the non-zero elements in AP matrix were abstracted by taking advantage of the sparsity features of partially-connected hybrid architecture to form an optimization vector. Simulation results indicate that our proposed method can achieve a better SR performance than existing methods such as MRT and Max-SR+NSP.


\ifCLASSOPTIONcaptionsoff
\newpage
\fi
\bibliographystyle{IEEEtran}
\bibliography{IEEEfull,REF}
\end{document}